\documentclass[a4paper,11pt]{article}

\usepackage[font=footnotesize]{caption}
\usepackage{array}
\usepackage{enumerate}
\usepackage{float}
\usepackage{subfig}
\usepackage{multicol}
\usepackage{multirow}
\usepackage{epsfig}
\usepackage{graphicx}
\usepackage{color}
\usepackage{cite}
\usepackage{amsmath,amsfonts,amssymb}
\usepackage{bbm}
\usepackage[utf8]{inputenc}
\usepackage[english]{babel}
\usepackage{xspace}
\usepackage{booktabs}

\usepackage[a4paper,top=2.5cm,bottom=2.5cm,left=2.5cm,right=2.5cm]{geometry}

\usepackage[pdftitle={Natural inflation and moduli stabilization in heterotic orbifolds},
   pdfauthor={Fabian Ruehle, Clemens Wieck},
   pdfsubject={},
   bookmarksopen, bookmarksnumbered, bookmarksopenlevel=2, colorlinks=false, linkcolor=blue, citecolor=blue, urlcolor=blue]{hyperref}

\newcommand{\U}[1]{\text{U(\ensuremath{#1})}\xspace}
\newcommand{\UA}[1]{\text{U(\ensuremath{#1})\ensuremath{_\text{A}}}\xspace}

\newcommand{\E}[1]{\ensuremath{\text{E}_{#1}}\xspace}

\renewcommand{\i}{\text{i}}
\newcommand{\Z}[1]{\ensuremath{\mathbb{Z}_{#1}}\xspace}

\begin{document}
\begin{flushright}
 DESY-15-040
\end{flushright}
\bigskip
\vspace{3cm}
\begin{center}
{\Large {\bf Natural inflation and moduli stabilization in heterotic orbifolds}
}
\\[0pt]
\bigskip {
{Fabian Ruehle\footnote{fabian.ruehle@desy.de}},
{Clemens Wieck\footnote{clemens.wieck@desy.de}}
\bigskip }\\[0pt]
\textit{Deutsches Elektronen-Synchrotron DESY, Notkestrasse 85, 22607 Hamburg, Germany}
\bigskip
\end{center}

\begin{abstract}
We study moduli stabilization in combination with inflation in heterotic orbifold compactifications in the light of a large Hubble scale and the favored tensor-to-scalar ratio $r \approx 0.05$. To account for a trans-Planckian field range we implement aligned natural inflation. Although there is only one universal axion in heterotic constructions, further axions from the geometric moduli can be used for alignment and inflation. We argue that such an alignment is rather generic on orbifolds, since all non-perturbative terms are determined by modular weights of the involved fields and the Dedekind $\eta$ function. We present two setups inspired by the mini-landscape models of the $\mathbb Z_{6-\text{II}}$ orbifold which realize aligned inflation and stabilization of the relevant moduli. One has a supersymmetric vacuum after inflation, while the other includes a gaugino condensate which breaks supersymmetry at a high scale.
\end{abstract}

\thispagestyle{empty}
\newpage
\setcounter{page}{1}
\tableofcontents

\section{Introduction}
\label{sec:Introduction}

Precision measurements of the CMB radiation increasingly favor the paradigm that the very early universe can be described by a phase of single-field slow-roll inflation \cite{Planck:2015xua,Ade:2015lrj}. In particular, recent observations of polarization fluctuations in the CMB indicate the possibility of substantial tensor modes among the primordial perturbations \cite{Ade:2014xna,Ade:2015tva}. This necessitates large-field models of inflation, i.e., the inflaton field must traverse a trans-Planckian field range during the last 60 $e$-folds of inflation \cite{Lyth:1996im}. Since large-field inflation is potentially susceptible to an infinite series of Planck-suppressed operators, this requires an understanding of possible quantum gravity effects. Thus, there has been renewed interest in obtaining inflation models from string theory. 

In this context natural inflation, first proposed in \cite{Freese:1990rb}, is among the most promising candidates. Here the flatness of the inflaton potential is guaranteed by an axionic shift symmetry which is exact in perturbation theory, but potentially broken by non-perturbative effects \cite{Wen:1985jz,Dine:1986vd}. Nevertheless, discrete symmetries may survive which protect the potential even at trans-Planckian field values. However, while axions are abundant in string theory compactifications we still face a problem: trans-Planckian inflaton values require an axion decay constant which is larger than the Planck scale. However, in string theory one generically expects the decay constant to be smaller than the string scale \cite{Banks:2003sx,Svrcek:2006yi}.

Different paths have been proposed to address this problem. In N-flation \cite{Dimopoulos:2005ac,Kim:2006ys}, for example, many axions with sub-Planckian decay constants contribute to the trans-Planckian field range of the inflaton, which is a linear combination of axions. However, this typically requires a very large number of axions which might be challenging to realize explicitly while maintaining control over the models. Another option was considered in \cite{Abe:2014pwa}, where the authors obtain trans-Planckian axions by choosing large gauge groups and by stabilizing the K\"ahler moduli at values much below the Planck scale. In that case, in principle one has to worry about perturbative control of the supergravity approximation, i.e., stringy corrections may be important. Furthermore, there is axion monodromy inflation \cite{Silverstein:2008sg,McAllister:2008hb} which uses a single sub-Planckian axion with a multi-valued potential to create an effectively trans-Planckian field range during inflation.

Another way of obtaining a large effective axion decay constant from a few number of axions is by alignment as proposed in \cite{Kim:2004rp} and further developed in \cite{Kappl:2014lra,Kappl:2015pxa}, or by kinetic alignment \cite{Bachlechner:2014hsa}.\footnote{See \cite{Tye:2014tja,Choi:2014rja} for related alignment mechanisms.} In the minimal setup of \cite{Kim:2004rp} there are two axions which appear as a linear combination in multiple non-perturbative contributions to the superpotential. If the axion decay constants are almost aligned one obtains an effective axion with a large decay constant, although the individual decay constants were small. In this paper we focus on the KNP alignment mechanism and its realization in $\E8\times\E8$ heterotic string theory \cite{Gross:1984dd} on orbifolds \cite{Dixon:1985jw,Dixon:1986jc}. Progress in this direction has recently been made in \cite{Ali:2014mra}, where the authors embedded aligned natural inflation in a supergravity model motivated by heterotic string compactifications on smooth Calabi-Yau manifolds with vector bundles. However, the authors did not specify the mechanism of moduli stabilization or an underlying reason for the alignment of the non-perturbative terms. The authors of \cite{Ben-Dayan:2014zsa} proposed a related model of hierarchical axion inflation and how it could be embedded in type IIB string theory. For other attempts to embed aligned natural inflation in type IIB string theory see \cite{Long:2014dta,Gao:2014uha,Abe:2014xja,Bachlechner:2014gfa,Shiu:2015uva,Shiu:2015xda}, and \cite{Grimm:2007hs} for a related analysis.

We study whether alignment of heterotic axions may be achieved by considering world-sheet instantons or a combination of the latter with gaugino condensates. Since the contributions arise from completely different mechanisms a natural question arises: why should the two effects be aligned? We attempt to answer this question, focusing our discussion on heterotic orbifolds where the moduli dependence of both effects, the condensing gauge group and the world-sheet instantons, can be computed using methods of conformal field theory. We argue that an alignment of the two terms is not as unnatural as one may think, essentially because the moduli dependence of both effects is determined by modular weights and Dedekind $\eta$ functions.

Furthermore, we address the issue of consistent moduli stabilization. Whenever inflation is discussed in string theory one desires a hierarchy of the form
\begin{align}\label{eq:hierarchy}
M_\text s, \, M_\text{KK} > M_\text{moduli} > H\,,
\end{align}
where $M_\text s$ denotes the string scale, $M_\text{KK}$ the Kaluza-Klein scale, and $H$ is the Hubble scale during inflation. This hierarchy is essential to ensure that inflation can be described by an effective four-dimensional supergravity theory where the inflaton is the only dynamical degree of freedom. In addition, in case of metastable vacua the barriers protecting the minima of the moduli must be larger than $H^2$. This is to avoid moduli destabilization during inflation as pointed out in \cite{Buchmuller:2004xr,Kallosh:2004yh}. Using the terms needed for successful inflation and other contributions to the superpotential we provide such a hierarchy explicitly for the complex dilaton field and the two K\"ahler moduli whose axions combine to form the effective inflaton. In a similar way this has recently been discussed for two K\"ahler moduli in aligned inflation in \cite{Kappl:2015pxa}.

While finalizing this project, concerns about large-field inflation models, among them aligned axion inflation, were raised by the authors of \cite{Rudelius:2015xta,Montero:2015ofa,Brown:2015iha}. They argue that most models involving trans-Planckian axions are generically challenged by potential contributions from gravitational instantons. In \cite{Montero:2015ofa} the authors outline examples involving Euclidean D-brane instantons in type IIB string theory. To fully assess the implications of these analyses it would be interesting to study the potentially dangerous gravitational instantons on heterotic orbifolds in order to explicitly check whether these instantons do arise or are forbidden by the orbifold symmetries. This is, however, left for future investigation.

This paper is organized as follows. In Section \ref{sec:Orbifolds} we briefly review important properties of orbifold spaces for reference in later sections. Afterwards, in Section \ref{sec:Inflation} we review the axion alignment mechanism and explain how to compute the various contributions to the superpotential. We discuss how this can be combined with moduli stabilization without spoiling the alignment or the dynamics of inflation. In Section \ref{sec:Example} we present two toy examples of our mechanism based on the $\Z{6-\text{II}}$ orbifold of the mini-landscape models \cite{Lebedev:2007hv,Lebedev:2008un}. Section \ref{sec:Conclusion} contains our conclusions and an outlook.

\section{Properties of orbifolds}
\label{sec:Orbifolds}
In this section we briefly review those properties of heterotic orbifolds relevant for our discussion. A good and detailed review can, for example, be found in \cite{Bailin:1999nk}. References \cite{Dundee:2010sb,Parameswaran:2010ec} discuss the relevance of these ingredients for moduli stabilization. 

In the construction of Abelian heterotic toroidal orbifolds one starts with a six-torus $T^6$ parameterized by three complex coordinates $z_{1,2,3}$ and mods out a discrete $\Z{N}$ symmetry\footnote{A similar discussion applies to the case of $\Z{M}\times\Z{N}$ orbifolds.} $\theta$, 
\begin{align}
 \theta:~(z_1,z_2,z_3)\mapsto(e^{2\pi\i n_1/N}z_1,e^{2\pi\i n_2/N}z_2,e^{2\pi\i n_3/N}z_3)=(e^{2\pi\i v_1}z_1,e^{2\pi\i v_2}z_2,e^{2\pi\i v_3}z_3)\;,
\end{align}
where we have defined the twist vector $v=(v_1,v_2,v_3)$. Requiring that the resulting singular space is Calabi-Yau imposes $v_1+v_2+v_3\in\Z{}$. The $\Z{6-\text{II}}$ orbifold, for example, has ${v=(1/3,-1/2,1/6)}$, i.e., it acts with an order-three rotation on the first, with an order-two rotation on the second, and with an order-six rotation on the third torus. Hence, each orbifold has $N$ twisted sectors $\theta^k$, $k=0,\ldots,N-1$. To ensure modular invariance of the one-loop string partition functions, these twists have to be accompanied by a shift in the $\E8\times\E8$ gauge degrees of freedom. This shift is parameterized by the shift-vector $V$. In addition, depending on the geometry one can allow for up to six independent Wilson lines $W_i$ on the torus.

The massless string spectrum is given in terms of the twist $v$, the shift $V$, and the Wilson lines $W_i$. In addition to the usual untwisted strings in the $\theta^0$ sector, which close already on the torus, it contains new string states in twisted sectors $\theta^k$ which are called twisted strings. These only close under the orbifold action and are thus forced to localize at orbifold fixed points. Depending on the orbifold action and the toroidal lattice, the amount of untwisted K\"ahler moduli $T_i$, $i=1,\ldots,h^{1,1}$ and complex structure moduli $U_j$, $j=1,\ldots,h^{2,1}$ may vary. In the $\Z{6-\text{II}}$ case, for example, one has $h^{1,1}=3$ and $h^{2,1}=1$. The $T_i$ parameterize the size of the three $T^2$ sub-tori while $U$ parameterizes the shape of the $T^2$ on which the orbifold has a $\Z2$ action.

\subsection{Modular transformations}
\label{subsec:ModularTrafos}
The K\"ahler and complex structure moduli have an $\text{SL}(2,\Z{})$ symmetry under which the $T_i$ transform as
\begin{align}
\label{eq:TDuality}
 T_i~\rightarrow~\frac{a_i T_i -\i \, b_i}{\i \, c_i T_i+d_i} \,,
\end{align}
and likewise for the moduli $U_j$. Here, $a_i, b_i, c_i, d_i \in \mathbb{Z}$ and $a_i d_i-b_i c_i =1$. 

At zeroth order the K\"ahler potential of the moduli reads
\begin{align}
  K_\text{moduli}=-\sum_{i=1}^{h^{1,1}}\ln(T_i+\overline{T}_i)-\sum_{j=1}^{h^{2,1}}\ln(U_j+\overline{U}_j)\,.
\end{align}
It is readily checked that under the transformation \eqref{eq:TDuality} the K\"ahler potential transforms as
\begin{align}
  K_\text{moduli} \rightarrow K_\text{moduli}+\sum_{i=1}^{h^{1,1}}\ln|\i\, c_iT^i+d_i|^2+\sum_{j=1}^{h^{2,1}}\ln|\i\, c_j U^j+d_j|^2\,.
\end{align}
Hence the shift symmetry of the moduli in the K\"ahler potential is protected by the modular symmetry. Since $G=K_\text{moduli} + K_\text{matter}  +\ln |W|^2$, which appears in the supergravity Lagrangian, has to be invariant we find that the superpotential has to transform with modular weight $-1$,
\begin{align}
  W \rightarrow W\,\prod_{i=1}^{h^{1,1}}(\i\, c_iT^i+d_i)^{-1}\,\prod_{j=1}^{h^{2,1}}(\i\, c_j U^j+d_j)^{-1}\,.
\end{align}
In addition to the K\"ahler and the superpotential, also the chiral fields have non-trivial modular transformations,
\begin{align}
  \Phi_\alpha~\rightarrow~\Phi_\alpha\,\prod_{i=1}^{h^{1,1}} \left(\i\,c_i T_i+d_i\right)^{m^i_\alpha}\,\prod_{j=1}^{h^{2,1}} \left(\i\,c_j U_j+d_j\right)^{\ell^j_\alpha}\,.
\end{align}
The modular weights $m^i_\alpha$ and $\ell^j_\alpha$ depend on the orbifold twisted sector $k$ and oscillator numbers. Defining $w_i(k)=kv_i\text{ mod }1$, they are given by \cite{Dixon:1989fj,Louis:1991vh,Ibanez:1992hc}
\begin{align}
\begin{alignedat}{1}
\label{eq:ModularWeights}
 m^i=\left\{
	    \begin{array}{ll}
		0\;,						\quad & \text{if}\: w_i=0\;,\\
		w_i-1-\widetilde{N}^i+\widetilde{N}^{i\,*}\;,	\quad & \text{if}\: w_i\neq0\;.
	    \end{array}
 \right.\\[4mm]
 \ell^j=\left\{
	    \begin{array}{ll}
		0\;,						\quad & \text{if}\: w_j=0\;,\\
		w_i-1+\widetilde{N}^j-\widetilde{N}^{j\,*}\;,	\quad & \text{if}\: w_j\neq0\;.
	    \end{array}
\right.
\end{alignedat}
\end{align}
Here, the $\widetilde{N}^i$ and $\widetilde{N}^{i\,*}$ are integer oscillation numbers. In the $p^\text{th}$ complex plane of the untwisted sector we have $m_p^i=-\delta_p^i$, $\ell_p^j=-\delta_p^j$. From this we find for the K\"ahler potential for the matter fields at lowest order 
\begin{align}\label{eq:Kmatter}
 K_\text{matter} = \sum_\alpha\prod_{i=1}^{h^{1,1}} \left(T_i + \overline{T_i}\right)^{m_\alpha^i}\,\prod_{j=1}^{h^{2,1}}\left(U_j + \overline{U}_j\right)^{\ell_\alpha^j}\, |\Phi_\alpha|^2\,.
\end{align}
Since the matter fields transform non-trivially and the superpotential has to have modular weight~$-1$, the coupling ``constants'' $y_{\alpha_1\ldots\alpha_L}$ of the $L$-point correlator 
\begin{align}
\label{eq:NPWSSchematic}
W\supset y_{\alpha_1\ldots\alpha_L}\Phi_{\alpha_1}\ldots\Phi_{\alpha_L}
\end{align}
have to be appropriate modular functions such that the overall modular weight is $-1$. Specifically,
\begin{align}
 y_{\alpha_1\ldots\alpha_L}\Phi_{\alpha_1}\ldots\Phi_{\alpha_L} \propto \prod_{i=1}^{h^{1,1}}~\prod_{j=1}^{h^{2,1}}\eta(T_i)^{2 r_i}~\eta(U_j)^{2 s_j}~\Phi_{\alpha_1}\ldots\Phi_{\alpha_L}\,,
\end{align}
where $\eta$ denotes the Dedekind $\eta$ function\footnote{In general, other modular functions can appear as well \cite{Lauer:1989ax,Ferrara:1989qb,Lauer:1990tm,Stieberger:1992bj}.} defined by
\begin{align}
 \eta(T)=e^{-\frac{\pi T}{12}}\prod_{\rho=1}^\infty \left(1-e^{-2\pi \rho T}\right)\,,
\end{align}
and the constant parameters $r_i$ and $s_j$ are determined by the modular weights,
 \begin{align}
 r_i=-1-\sum_\alpha m_\alpha^i\,,\qquad s_j=-1-\sum_\alpha \ell_\alpha^j\,.
\end{align}
The Dedekind $\eta$ function transforms under modular transformations up to a phase, 
\begin{align}
 \eta(T)\rightarrow (\i\, c T+d)^{1/2}~\eta(T)\,.
\end{align}
For $T > 1$ in Planck units we use the approximation
\begin{align}
 \label{eq:EtaApprox}
 \eta(T)=e^{-\frac{\pi T}{12}}\,.
\end{align}
As a result the non-perturbative superpotential terms are of the schematic form
\begin{align}
\label{eq:NPWS}
 W_\text{NP}^{\text{WS}}=A(\Phi_\alpha)\;e^{-\frac{2\pi}{12}\left(\sum_i r_i T_i+\sum_j s_j U_j\right)}\,.
\end{align}
Note that, if the fields $\Phi_\alpha$ are charged under an anomalous \U1 symmetry, $S$ may appear in the exponent as well. In particular, this is the case when the model-independent axion contained in $S$ cancels the anomalies, as explained in more detail below.

\subsection{Anomalous U(1) and FI terms}
\label{subsec:AnomalousU1}
In orbifold models with shift embeddings, the primordial $\E8\times\E8$ gauge symmetry is broken rank-preservingly into Abelian and non-Abelian gauge factors. Generically one \U1 is anomalous, henceforth denoted by \UA1. This anomaly is canceled via a Green-Schwarz (GS) mechanism \cite{Green:1984sg}. More precisely, the dilaton $S$ transforms under such an anomalous gauge variation as $S\rightarrow S-\i \Lambda \delta_\text{GS}$, where $\Lambda$ is the superfield gauge parameter and $\delta_\text{GS}$ is a real constant. As a consequence the combination $S+\overline{S}-\delta_\text{GS}V_A$ is gauge-invariant, where $V_A$ is the vector multiplet associated with \UA1. 

The non-trivial \UA1 transformation of $S$ has two important consequences. First, we observe that GS anomaly cancellation results in a field-dependent Fayet-Iliopoulos (FI) term\footnote{Notice that this commonly used terminology is slightly misleading. A field-dependent FI term is usually the $D$-term of a complex field with a logarithmic K\"ahler potential, which, if integrated out at a high scale, may mimic a constant FI term as the one introduced in \cite{Fayet:1974jb}. We refer to the original discussion in \cite{Dine:1987xk} for more details.} of the form
\begin{align}
 \label{eq:FITerm}
 \xi=\frac{\delta_\text{GS}}{(S+\overline{S})}\;.
\end{align}
In order to preserve $D$-flatness, this means that some chiral orbifold fields $\Phi_\alpha$ with appropriate charge must get a vacuum expectation value (VEV) to cancel $\xi$. The VEV of these fields can, at the same time, break unwanted extra gauge groups and lift vector-like exotics and other extra hidden fields in a Higgs-like mechanism. Generically, the primordial $\E8\times\E8$ is broken to many U(1) factors under which the orbifold fields are charged simultaneously. Hence, $D$-flatness of the other U(1) symmetries requires that many fields obtain a non-vanishing VEV. Second, superpotential terms involving the dilaton in the exponent have to be such that the whole correlator is gauge-invariant.

Moreover, $S$ has a non-trivial modular transformation to ensure anomaly cancellation in the underlying sigma-model \cite{Dixon:1990pc,Derendinger:1991hq}:
\begin{align}
\label{eq:DilatonTrafo}
S\rightarrow S+\frac{1}{8\pi^2}\sum_{i=1}^{h^{1,1}}\delta^i\ln(\i c_i T_i+d_i)+\sum_{j=1}^{h^{2,1}}\delta^j\ln(\i c_j U_j+d_j) \,,
\end{align}
where $\delta^i$ and $\delta^j$ are real constants of order 1 that can be computed from the sigma-model anomaly cancellation condition. As a consequence, the modular invariant K\"ahler potential of the dilaton reads 
\begin{align}
\label{eq:KDilatonExact}
 K_\text{dilaton}=-\ln(Y)=-\ln\left(S+\overline{S}+\frac{1}{8\pi^2}\left[\sum_{i=1}^{h^{1,1}}\delta^i\ln(T_i+\overline{T}_i)+\sum_{j=1}^{h^{2,1}}\delta^j\ln(U_j+\overline{U}_j)\right]\right)\,.
\end{align}
Due to the loop suppression factor $8 \pi^2$, these corrections are small as long as the $T_i$ are not stabilized at substantially larger field values than $S$. This is not the case in the models we study.

\subsection{Gauge kinetic function and gaugino condensation}
\label{subsec:GaugeKinFunction}
The one-loop gauge kinetic function of a gauge group $G_a$ at Ka\v{c}--Moody level 1 is given by \cite{Dixon:1989fj,Dixon:1990pc,Lust:1991yi}
\begin{align}\label{eq:GKF}
 f_a(S,T,U)=S+\frac{1}{8\pi^2}\sum_{i=1}^{h^{1,1}}b_a^i(m) \frac{g_i}{N}\ln(\eta(T^i))^2+\frac{1}{8\pi^2}\sum_{j=1}^{h^{2,1}}b_a^j(\ell) \frac{g_j}{N}\ln(\eta(U^j))^2\;,
\end{align}
where $b_a^i$ are the $\beta$-function coefficients in the $i^\text{th}$ torus of the gauge group $G_a$. They are non-vanishing in the $\mathcal{N}=2$ twisted sub-sectors of the theory and depend on the Dynkin indices and on the modular weights of the states charged under $G_a$. Furthermore, the $g_i$ are the order of the little group of the orbifold action in the $i^\text{th}$ torus, i.e., the order of the group that leaves the  $i^\text{th}$ torus fixed.  Depending on the lattice and the presence of Wilson lines, the modular symmetry group SL(2,\Z{}) might be reduced such that only a subgroup $\Gamma^0(N/g_i)$ or $\Gamma_0(N/g_i)$ is realized \cite{Bailin:1993fm,Bailin:1993ri,Bailin:2014nna}. In the example of the factorized $\Z{6-\text{II}}$ orbifold the $\mathcal{N}=2$ twisted sectors are $\theta^k$ with $k=2,3,4$, $N=6$, $g_1=2$, $g_2=3$ and the modular group is not reduced.

The gauginos of $G_a$ may condense at a scale $\Lambda_a^\text{GC}$ which depends on the low-energy effective $\mathcal{N}=1$ $\beta$-function, given by
\begin{align}
\label{eq:EffectiveBetaFunction}
 \beta_a = \frac{11}{3}C_2(\textbf{Ad}_a)-\frac{2}{3}\left(C_2(\textbf{Ad}_a)+\sum_{\psi_{\mathbf{R}_a}}C_2(\mathbf{R}_a)\right)-\frac{1}{3}\sum_{\phi_{\mathbf{R}_a}}C_2(\mathbf{R}_a)\,,
\end{align}
where $C_2(\mathbf{R}_a)$ is the quadratic Casimir operator of the irreducible representation $\mathbf{R}_a$. $\Lambda_a^\text{GC}$ can then be written in terms of the gauge kinetic function as \cite{Taylor:1982bp,Affleck:1983mk}, 
\begin{align}
\label{eq:CondensationScale}
 \Lambda_a^\text{GC} = e^{-\frac{8\pi^2}{\beta_a}f_a(S,T,U)}\,.
\end{align}
As discussed above, all extra fields become massive. If their mass is larger than the condensation scale they can be integrated out. The $\beta$-function \eqref{eq:EffectiveBetaFunction} is then simply $3\check{c}$, where $\check{c}$ denotes the dual Coxeter number. The effective superpotential term generated by gaugino condensation is $\propto (\Lambda_a^\text{GC})^3$. In addition, it depends on the fields charged under the condensing gauge group and on the fields that get a VEV and give an effective mass term to those fields. The final expression involves, in addition to the $\mathcal{N}=2$ beta function of the condensing gauge group, the modular weights of the fields that enter in the condensate. To obtain the final expression, we insert \eqref{eq:GKF} into \eqref{eq:CondensationScale}, and include a field-dependent pre-factor from integrating out the heavy fields \cite{Binetruy:1996uv}. Using the transformation behavior of the dilaton \eqref{eq:DilatonTrafo} and requiring that the result has again modular weight $-1$, we find
\begin{align}
\label{eq:NPGC}
 W_\text{NP}^{\text{GC}}=B(\Phi_\rho)\;e^{-\frac{8\pi^2}{\check{c}}S+\sum_{i}(-2+\frac{2\delta^i}{\check{c}})\ln\eta(T_i)+\sum_{j}(-2+\frac{2\delta^j}{\check{c}})\ln\eta(U_j)}\,.
\end{align}
Hence, we observe that both the non-perturbative world-sheet instanton contributions \eqref{eq:NPWS} and the non-perturbative gaugino condensation terms \eqref{eq:NPGC} depend on the modular weights and on the Dedekind $\eta$ function. The combined superpotential, using \eqref{eq:GKF} and \eqref{eq:EtaApprox}, has the schematic form
\begin{align}
\label{eq:SPotFull}
 W \supset \prod_\alpha \Phi_\alpha\;e^{-\sum_\alpha \frac{q_\alpha}{\delta_\text{GS}} Y - \frac{2\pi}{12}\left( \sum_{i} r_i T_i + \sum_{j} s_j U_j\right)}+ B(\Phi_\rho)e^{-\frac{8\pi^2}{\check{c}}S+\frac{2\pi}{12}(\sum_{i} b_i T_i+\sum_{j} b_j U_j)}\,,
\end{align}
where $q_\alpha$ are the \UA1 charges of the fields $\Phi_\alpha$. Note that the modular weights $m$ and $\ell$ are negative and such that $r_i,s_j\geq0$ for most couplings. The constants $b_i$ and $b_j$ also depend on the modular weights and in addition on the $\mathcal{N}=2$ beta function coefficients,
\begin{align}
 b_i=1-\sum_i \frac{\delta^i}{\check{c}}\,,\qquad b_j=1-\sum_j \frac{\delta^j}{\check{c}}\,.
\end{align}
As mentioned before, the $\delta^i$ and $\delta^j$ are typically of order 1 so that $b_i, b_j\approx1$, especially for large gauge groups. Note that in many couplings at least some of these constants are zero and hence the corresponding modulus does not appear in those superpotential terms.

\section{Inflation in heterotic orbifolds}
\label{sec:Inflation}

Let us now discuss how inflation can be realized in heterotic orbifold compactifications. We briefly review the alignment mechanism proposed in \cite{Kim:2004rp,Kappl:2014lra} and subsequently put the ingredients of Section \ref{sec:Orbifolds} together to build an aligned axion inflation model with all moduli stabilized at a high scale.

\subsection{The alignment mechanism}
\label{subsec:AlignmentMechanism}
Remember that alignment means, on the level of the effective potential for two axions $\tau_{1,2}$,
\begin{align}\label{eq:effpot}
 V=\kappa_1\left(1-\cos(\beta_1\tau_1+\beta_2\tau_2)\right)+\kappa_2\left(1-\cos(n_1\tau_1+n_2\tau_2)\right)\,,
\end{align}
that there is a flat direction if
\begin{align}
\label{eq:ratios}
 \frac{\beta_1}{n_1}=\frac{\beta_2}{n_2}\,.
\end{align}
Notice that the coefficients $\beta_i$ and $n_i$ are the inverse of the axion decay constants. To slightly lift this flat direction one can introduce a small misalignment parameterized by \cite{Kappl:2014lra}
\begin{align}
\label{eq:AlignmentParameter}
 k:=\frac{1}{n_2}-\frac{\beta_1}{\beta_2}\frac{1}{n_1}\,,
\end{align}
which vanishes for perfect alignment. After rotating to a convenient field basis, ${(\tau_1,\tau_2)\mapsto(\varphi_1,\varphi_2)}$ and canonically normalizing the kinetic terms, we obtain for the almost flat direction $\varphi_1$ an effective decay constant $f_\text{eff}$ which reads \cite{Kappl:2014lra,Ali:2014mra}
\begin{align}
 f_\text{eff}=\frac{\beta_1^2 \sqrt{(\beta_1^{-2}+\beta_2^{-2})(\beta_1^{-2}+n_1^{-2})}}{k n_1 \beta_2}\,.
\end{align}
It is arbitrarily large for arbitrarily small $k$ and hence closely aligned axions $\tau_i$. A sizeable tensor-to-scalar ratio $r\approx0.05$ requires a misalignment of $k\approx0.2$. 

\subsection{Alignment and moduli stabilization on orbifolds}
\label{subsec:ModuliStabilization}
A complete treatment of stabilizing all moduli while keeping three MSSM generations of particles and one pair of Higgs fields with realistic Yukawa couplings, decoupling extra vector-like exotics, and breaking additional U(1) symmetries generically present in these models is beyond the scope of this paper. Moduli stabilization in similar setups without considering inflation has been investigated in \cite{Dundee:2010sb,Parameswaran:2010ec}. However, the mechanisms used there typically yield masses below the currently favored large Hubble scale and are thus incompatible with single-field inflation.

From the discussion in Section \ref{sec:Orbifolds} it should be clear that the effective potential \eqref{eq:effpot} is sourced by a superpotential with two non-perturbative terms, both of which contain two K\"ahler moduli $T_1$ and $T_2$. In particular, we mostly focus on the two K\"ahler moduli which correspond to the tori that have an $\mathcal{N}=2$ sub-sector.\footnote{In the prime orbifolds $\Z3$ and $\Z7$ no torus has an $\mathcal{N}=2$ sub-sector while in the $\Z{M}\times\Z{N}$ orbifolds all three tori do.} In fact, all orbifolds have at least three untwisted K\"ahler moduli and up to three untwisted complex structure moduli. Concerning their stabilization, note that those K\"ahler moduli which correspond to tori that have fixed points in all twisted sectors $\theta^k$ do not enter in the gauge kinetic function and thus can only be stabilized via world-sheet instantons. Whether they appear in a world-sheet instanton coupling depends on the modular weights as discussed above. For the sake of simplicity we assume that the moduli not involved in the stabilization or alignment mechanism, as well as other potentially present fields, have been stabilized at a scale above $H$ and consequently decouple from inflation. 

The real parts of the $T_i$ govern the size of the compactification manifold. The imaginary parts, albeit not involved in the anomaly cancellation except for the small one-loop contribution, enjoy an axionic shift symmetry inherited from the SL(2,\Z{}) symmetry. They yield a cosine-potential as in \eqref{eq:effpot} and can consequently be used as inflaton candidates. The real part of the complex dilaton field determines the gauge coupling strength while its imaginary part is the so-called universal axion which is responsible for Green-Schwarz anomaly cancellation, cf.~Section~\ref{subsec:AnomalousU1}. For a suitable choice of the superpotential, comprised of the terms generically available in orbifold compactifications, the effective potential after integrating out all moduli and the universal axion takes the form \eqref{eq:effpot}. In the following we discuss which parts of the superpotential may achieve this while ensuring consistent stabilization of the aforementioned relevant moduli.

\subsubsection*{Inflation with world-sheet instantons only}
A first option is to employ only world-sheet instanton contributions. For two aligned K\"ahler moduli this has recently been discussed in \cite{Kappl:2015pxa}, based on the mechanism proposed in \cite{Wieck:2014xxa}. We extend this to include dilaton stabilization by considering the part of the orbifold superpotential which has the form \begin{align}
\label{eq:WWSInstantons}
\begin{split}
W &= \chi_1 \left[A_1(\phi_\alpha,\chi_\beta) e^{-n_1 T_1 - n_2 T_2 } - P_1(\chi_\gamma) \right] + \chi_2 \left[ A_2(\phi_\mu,\chi_\nu) e^{-n_3 T_1 - n_4 T_2 } - P_2(\chi_\rho) \right] \\
&+ \chi_3 \left[A_3(\chi_\sigma) e^{-\frac{q}{\delta_\text{GS}} S} - P_3(\chi_\lambda) \right]\,,
\end{split}
\end{align}
where the $\chi_i$ and $\phi_i$ are untwisted and twisted chiral superfields, respectively, and $n_i=\frac{\pi}{6}r_i$ in the notation of \eqref{eq:SPotFull}. Note that we have neglected the loop contributions to $S$. To obtain the correct $T_i$ dependence in the various terms twisted fields necessarily enter the functions $A_i$, while the $P_i$ are functions of untwisted moduli. Since untwisted fields have modular weight $-1$ they do not induce a moduli dependence of the couplings. Likewise, the moduli dependence in the couplings of $A_1$ and $A_2$ arises from the twisted fields. The discussion has again been tailored to the $\Z{6-\text{II}}$ orbifold. For other orbifolds, especially for $\Z{M}\times\Z{N}$ orbifolds, there also exist couplings that involve only twisted states which nevertheless have modular weight $-1$, so that no extra $T_i$ occur in these terms.

In the above parameterization we assume that the fields entering $A_i$ and $P_i$ obtain non-vanishing vacuum expectation values via $F$- and $D$-terms. In our supergravity analysis we treat them as numerical constants given by the VEVs of these fields. Those VEVs are generically of the order of the string scale, $M_\text s \lesssim 0.1$. We assume that the other fields we have not made explicit obtain a mass in a similar way from couplings to fields that get a VEV.

The effective theory defined by \eqref{eq:WWSInstantons} and the K\"ahler potential discussed in Section \ref{sec:Orbifolds} has a supersymmetric Minkowski vacuum at $\langle \chi_1 \rangle = \langle \chi_2 \rangle = \langle \chi_3 \rangle = 0$. The auxiliary fields of the $\chi_i$ stabilize the complex scalars $S$, $T_1$, and $T_2$ at mass scales determined by the $A_i$ and $P_i$. In the heterotic mini-landscape models of \cite{Lebedev:2007hv,Lebedev:2008un} there are many examples in which the coefficients $n_i$ are such that sufficient alignment is possible. There is then a light linear combination of $T_1$ and $T_2$ whose imaginary part is the inflaton field. All other degrees of freedom can be sufficiently stabilized in many examples. More details and an explicit example which realizes the hierarchy~\eqref{eq:hierarchy} are given in Section \ref{sec:Example}.

After inflation has ended supersymmetry must be broken to avoid phenomenological problems. As pointed out in \cite{Kappl:2015pxa} the above scheme can accommodate low-energy supersymmetry breaking, for example, via the $F$-term of a Polonyi field. A more generic situation on orbifolds is supersymmetry breaking via gaugino condensates. As is well-known, these can also lead to a suppression of the supersymmetry-breaking scale compared to the Hubble scale. From the perspective of aligned inflation this is desirable in the above setup, since the gaugino condensate must not interfere with the alignment of $T_1$ and $T_2$. No matter how supersymmetry is broken, the resulting vacuum will have a positive cosmological constant which is determined by the scale of supersymmetry breaking. This must be cancelled by a fine-tuned constant contribution to the superpotential to high accuracy.

Since many non-Abelian gauge groups arise from the breaking of $\E8\times\E8$ the appearance of gaugino condensates is quite generic in orbifolds. In the following we discuss an example in which a gaugino condensate participates in the alignment mechanism.

\subsubsection*{Inflation with world-sheet instantons and gaugino condensates}
A second option to achieve alignment and moduli stabilization is to use a combination of gaugino condensates and world-sheet instantons. In this case supersymmetry is necessarily broken at a high scale in order to stabilize all fields above the Hubble scale. In many models we find superpotentials of the form \eqref{eq:SPotFull}, or more specifically
\begin{align}
\label{eq:WGCAndWSInstantons}
W = \chi_1 \left[B_1 e^{-\frac{8 \pi^2}{\check c} S + \beta_1 T_1 + \beta_2 T_2 } - P_1 \right] + \chi_2 \left[A_1  e^{- n_1 T_1 - n_2 T_2} - P_2 \right]+ \chi_3 \left[A_2 e^{-\frac{q}{\delta_\text{GS}} S} - P_3 \right]\,,
\end{align}
with $\beta_i = \frac{\pi}{6}b_i$. The notation and the field dependence of the $A_i$, $P_i$, $n_i$ is as in the previous example, and again we assume them to be constants arising from other fields that obtain a VEV. As explained in Section \ref{subsec:GaugeKinFunction} the $\beta_i$ depend on the particle content of the $\mathcal{N}=2$ sub-sector and the modular weights of $\chi_1$ and the fields entering $B_1$. In the effective theory of inflation $B_1$ is assumed to be constant as well. It is in general a non-analytic function of mesonic degrees of freedom which are integrated out above the scale of gaugino condensation. As explained in more detail in \cite{Dundee:2010sb}, $B1 \langle \chi_1 \rangle$ determines the meson mass in the vacuum, which must be larger than $H$ and the condensation scale to ensure decoupling in the effective theory and during inflation. This means that $\langle \chi_1 \rangle \neq 0$ in such setups. This is typically guaranteed by $D$-terms associated with \UA1 or, as in the above case, other U(1) symmetries. This means that the superpotential in \eqref{eq:WGCAndWSInstantons} yields a type of racetrack potential for the moduli, which are stabilized by their own $F$-terms and those of the $\chi_i$.

Moduli stabilization via the superpotential in \eqref{eq:WGCAndWSInstantons} generically yields dS vacua since supersymmetry is broken by the gaugino condensate. The scale of supersymmetry breaking is proportional to $B_1 \langle \chi_1 \rangle$ and necessarily lies, as explained above, close to the inflationary Hubble scale. However, to avoid a potentially destructive back-reaction of the auxiliary fields responsible for supersymmetry breaking, cf.~the discussion in \cite{Buchmuller:2014pla}, one must find examples in which the gravitino mass is not substantially larger than $H$. We demonstrate that this is possible in a second benchmark model in Section \ref{sec:Example}.

\section{Two benchmark models}
\label{sec:Example}
Let us now turn to two examples. We chose to discuss inflation in the context of the $\Z{6-\text{II}}$ mini-landscape models because these are the most-discussed models in the literature. However, the mechanisms discussed here apply to most orbifolds in a similar vein.

\subsection{Example 1: World-sheet instantons only}
Let us start with the situation described in Section \ref{subsec:ModuliStabilization}, where we stabilize the moduli via world-sheet instantons only. We assume that some untwisted and twisted fields $\Phi_\alpha$ have obtained a string-scale VEV from $D$-terms which we do not include explicitly here. As explained above, we take the $\chi_i$ to be untwisted and the $\phi_i$ to be twisted matter fields. Furthermore, we consider $\chi_3$ to carry \UA1 charge $q=1$.

The K\"ahler potential in this case reads
\begin{align}
K = -\ln{\left( S + \overline S \right)} -\ln{\left( T_1 + \overline T_1 - |\chi_1|^2 \right)} -\ln{\left( T_2 + \overline T_2 - |\chi_2|^2 \right)} + |\chi_3|^2\,,
\end{align}
where we have neglected the loop contributions to $S$. The contributions \eqref{eq:Kmatter} of the twisted fields do not affect the results of our discussion as long as all VEVs of the matter fields are of the order of the string scale or below.
We consider the part of the full superpotential given in~\eqref{eq:WWSInstantons}. The possible values for the modular weights $n_i$ are taken from the \texttt{orbifolder} \cite{Nilles:2011aj}, in this case $n_1 = \pi/6$, $n_2 = \pi/6$, $n_3 = \pi/3$, and $n_4 = \pi/2$. In typical models $\delta_\text{GS}\sim\mathcal{O}(0.1)$ and the \UA1 charges of the fields entering $A_2$ are $\mathcal{O}(1)$, such that we obtain an overall prefactor of $S$ of order $1$. The VEVs of the fields entering $A_2$ cancel the $D$-term induced by $\delta_\text{GS}$. The remaining input parameters for this example are summarized in Table~\ref{tab:Ex1Params}.
\begin{table}[t]
 \centering
 \begin{tabular}{cccccc}
 \toprule
 $A_1$ & $A_2$ & $A_3$ & $P_1$ & $P_2$ & $P_3$ \\
 \midrule
 $3.2 \cdot 10^{-4}$ & $6.8 \cdot 10^{-4}$ & $1.6 \cdot 10^{-3}$ & $9.7 \cdot 10^{-5}$ & $3.2 \cdot10^{-5}$ & $2.6 \cdot 10^{-4}$	\\
 \bottomrule
 \end{tabular}
 \caption{Input parameters for the constants used in Example 1. The $A_i$ and $P_i$ arise from $3$- and $4$-point couplings. \label{tab:Ex1Params}}
\end{table}
The resulting theory has a supersymmetric vacuum at $\langle \chi_1 \rangle = \langle \chi_2 \rangle = \langle \chi_3 \rangle = \langle \text{Im}\, S \rangle = 0$ and $\langle \text{Re}\, S \rangle \approx 1.8$. The lightest eigenvalue in the mass matrix corresponds to the aligned linear combination of $T_1$ and $T_2$. A convenient field basis is therefore
\begin{align}
T_1 \to \tilde T_1 =  a T_1 + b T_2\,, \qquad
T_2 \to \tilde T_2 = -b T_1 + a T_2\,,
\end{align}
with $a \approx -0.64$ and $b \approx -0.77$ in this case. $\tilde T_2$ is the lightest direction and $\text{Im} \, \tilde T_2$ is the inflaton. In the vacuum its real part is as heavy as the inflaton because supersymmetry is unbroken. Thus, one may worry that it contributes quantum fluctuations to the system, yielding a multi-field inflation model. However, during inflation $\text{Re} \, \tilde T_2$ receives a soft mass term of the same order as the Hubble scale. Indeed, a numerical analysis of the coupled equations of motion, similar to the one carried out in \cite{Kappl:2015pxa}, reveals that all fields except the inflaton are sufficiently stabilized during inflation. For 60 $e$-folds of slow-roll inflation we summarize the predictions for the CMB observables and other relevant parameters in Table \ref{tab:Ex1CMB}.
\begin{table}[t]
 \centering
 \begin{tabular}{ccccc}
 \toprule
 $\langle T_1 \rangle = \langle \overline T_1 \rangle$ & $\langle T_2 \rangle = \langle \overline T_2 \rangle$ & $f_\text{eff}$ & $n_\text s$ & $r$  \\
 \midrule
 $ 1.06 $ & $ 1.24 $ & $ 5.7 $ & $ 0.96 $ & $ 0.03 $\\
 \bottomrule
 \end{tabular}
 \caption{CMB observables and other relevant parameters for 60 $e$-folds of inflation in Example 1. \label{tab:Ex1CMB}}
\end{table}

Apparently, successful inflation in line with recent observations is possible in this setup. However, since we have chosen to only employ a portion of the total superpotential of such orbifold models, one may worry about additional terms which can interfere with moduli stabilization or the alignment mechanism. In particular, there may be terms of the form 
\begin{align}
W \supset C(\Phi_\alpha) e^{-f(T_1,T_2)}\,,
\end{align}
where the function $f$ contains some linear combination of the two moduli. On the one hand, this term clearly breaks supersymmetry if $ C(\Phi_\alpha) \neq 0$. The effects on inflation, however, are not significant as long as the resulting gravitino mass is not much larger than $H$, which is generically fulfilled. On the other hand, the additional dependence on the moduli may interfere with the alignment of the effective inflaton field. We have verified that this is negligible as long as $C < A_i$. This means that additional terms of this type must be suppressed up to slightly higher order than the ones in the part of the superpotential we consider.

\subsection{Example 2:  World-sheet instantons and gaugino condensates}

The setup which includes a gaugino condensate is slightly more generic, but also more complicated. Similar to the previous example the K\"ahler potential reads
\begin{align}
K = -\ln{\left( S + \overline S \right)} -\ln{\left( T_1 + \overline T_1 - |\chi_1|^2 \right)} -\ln{\left( T_2 + \overline T_2 - |\chi_2|^2 \right)} + |\chi_3|^2\,,
\end{align}
where we have once more neglected the loop-suppressed correction to the dilaton K\"ahler potential and the contribution of the twisted matter fields. The superpotential is this time given by \eqref{eq:WGCAndWSInstantons} with $n_1 = \pi/2$, $n_2 = \pi/3$, $\beta_1 = \pi/6$, $\beta_2 = \pi/6$ and $q/\delta_\text{GS}=1$. As in the previous example, the FI term of \UA1 is canceled by the VEVs of the fields entering $A_2$. Note that $\chi_3$ cannot cancel this FI term since we assume in our example that its charge has the wrong sign. Nevertheless, on orbifolds fields are typically charged under many U(1) factors simultaneously. To account for this, we include a $D$-term $\zeta$ originating from another U(1) under which $\chi_{3}$ has charge $-1$, 
\begin{align}
V_D = \frac{1}{S+ \overline S}\left( \chi_3 K_{\chi_3} -\zeta \right)^2\,,
\end{align}
with $\zeta = 10^{-3}$. This $D$-term is canceled by $\langle \chi_3\rangle\neq0$, which results in a non-vanishing VEV of the other fields, $\langle \chi_{1,2} \rangle \neq 0$. All other input parameters are summarized in Table~\ref{tab:Ex2Params}.
\begin{table}[t]
 \centering
 \begin{tabular}{cccccc}
 \toprule
 $B_1$ & $A_1$ & $A_2$ & $P_1$ & $P_2$ & $P_3$ \\
 \midrule
 $ 14.4 $ & $ 6.8 \cdot 10^{-3} $ & $ 3.6 \cdot 10^{-3} $ & $ 2.1 \cdot 10^{-4} $ & $ 9.1 \cdot 10^{-5} $ & $ 7.1 \cdot 10^{-4} $ \\
 \bottomrule
 \end{tabular}
 \caption{Input parameters for the constants used in Example 2. The $A_i$ and $P_i$ arise from $3$- and $4$-point couplings.  \label{tab:Ex2Params}}
\end{table}
The resulting scalar potential has a dS vacuum specified in Table~\ref{tab:Ex2VEVs}. 
\begin{table}[t]
 \centering
 \begin{tabular}{cccccc}
 \toprule
$ \langle S \rangle = \langle \overline S \rangle $ & $\langle T_1 \rangle = \langle \overline T_1 \rangle$ & $\langle T_2 \rangle = \langle \overline T_2 \rangle$ & $\langle \chi_1 \rangle $ & $ \langle \chi_2 \rangle $ & $ \langle \chi_3 \rangle $  \\
 \midrule
 $ 1.6 $ & $ 1.97 $ & $ 1.16 $ & $ 9.6 \cdot 10^{-3} $ & $ -7.9 \cdot 10^{-2} $ & $ -2.2 \cdot 10^{-2} $ \\
 \bottomrule
 \end{tabular}
 \caption{Vacuum expectation values of all relevant fields in the dS minimum with $\langle V \rangle \approx 2 \cdot 10^{-14}$. In addition, the imaginary parts of the $\chi_i$ obtain VEVs much smaller than 1. \label{tab:Ex2VEVs}}
\end{table}
The positive vacuum energy can be cancelled by a fine-tuned constant contribution to $W$, and the gravitino mass in the near-Minkowski vacuum is $m_{3/2} \approx 6.2 \cdot 10^{-7}$. There is again a lightest direction in the mass matrix which is $\tilde T_2$ with $a \approx -0.82$ and $b \approx -0.56$, and its imaginary part is the inflaton. Once more we solve the coupled equations of motion to ensure that all other degrees of freedom are sufficiently stable during inflation. The CMB predictions for this second case are summarized in Table \ref{tab:Ex2CMB}.
\begin{table}[t]
 \centering
 \begin{tabular}{ccc}
 \toprule
$f_\text{eff}$ & $n_\text s$ & $r$  \\
 \midrule
 $ 5.7 $ & $ 0.96 $ & $ 0.04 $\\
 \bottomrule
 \end{tabular}
 \caption{CMB observables for 60 $e$-folds of inflation and the effective axion decay constant in Example 2. \label{tab:Ex2CMB}}
\end{table}
Further contributions to the superpotential must satisfy the same constraints as in Example 1 to not interfere with inflation.

\section{Conclusions}
\label{sec:Conclusion}

We have analyzed the feasibility of natural inflation with consistent moduli stabilization in heterotic orbifold compactifications. To allow for the trans-Planckian axion field range favored by recent observations of the CMB polarization, we implement aligned natural inflation with two axions. Generic properties of orbifolds naturally permit sufficient alignment for 60 $e$-folds of slow-roll inflation with a detectable tensor-to-scalar ratio and a scalar spectral index of $n_\text s \approx 0.96$, and at the same time provide a mechanism to stabilize the relevant moduli and the dilaton.

The alignment is produced by two non-perturbative terms in the superpotential of the orbifold. They may either be sourced by two world-sheet instantons which couple to twisted and untwisted matter fields, or by a world-sheet instanton and a gaugino condensate of a non-Abelian gauge group in the hidden sector of the primordial $\E8 \times \E8$. The axions which mix are the imaginary parts of two complex untwisted K\"ahler moduli, governing the size of two tori. A crucial observation is that both possible non-perturbative effects are determined by the modular weights of the fields involved in the coupling and the Dedekind $\eta$ function. This leads to many instantonic couplings with similar coefficients in the exponential, corresponding to the individual axion decay constants, which in turn allows for aligned inflation. Since any embedding of inflation in string theory must address moduli stabilization, we demonstrate how both K\"ahler moduli and the dilaton can be stabilized at a high scale. This can happen through the terms needed for inflation and additional terms involving the VEVs of twisted and untwisted matter fields.

In the case of two world-sheet instantons all axion coefficients are determined by sums of modular weights and the Dedekind $\eta$ function. Thus, the more fields are involved in the correlator, the larger the coefficients of the moduli in the instantonic terms. This way, couplings generated at fourth or higher order generically have coefficients which allow for just the right amount of alignment. The case in which inflation is driven by a world-sheet instanton and a gaugino condensate is more constrained, and thus more predictive. The coefficients in the gaugino condensate are fixed by symmetry arguments and the Dedekind $\eta$ function. Alignment can occur when the world-sheet instanton coupling is introduced at sufficiently high order. In both cases, additional terms in the superpotential do not interfere with inflation or moduli stabilization, as long as their magnitude is below the inflationary Hubble scale.

We provide benchmark models for both cases to illustrate our findings. In the first case we find a supersymmetric Minkowski vacuum in which the flattest direction is a linear combination of the two K\"ahler moduli, the imaginary part of which is the aligned inflaton. All other degrees of freedom are stabilized at a higher scale and decouple from inflation. During inflation the real part of the aligned modulus receives a Hubble-scale soft mass and is sufficiently stable as well. This situation is similar in the second case, although in the vacuum supersymmetry is spontaneously broken by the gaugino condensate with $m_{3/2} \lesssim H$.

\section*{Acknowledgments}
We thank Wilfried Buchm\"uller, Emilian Dudas, Rolf Kappl, Hans Peter Nilles, Martin Winkler, and Alexander Westphal for useful discussions. This work was supported by the German Science Foundation (DFG) within the Collaborative Research Center (SFB) 676 ``Particles,
Strings and the Early Universe''. The work of C.W.~has been supported by a scholarship of the Joachim Herz Foundation.

\begin{footnotesize}

\providecommand{\href}[2]{#2}
\end{footnotesize}
\end{document}